\documentclass[aps,prc,twocolumn,groupedaddress,showpacs,superscriptaddress,nofootinbib]{revtex4-1}
\usepackage{graphicx,amsmath}
\usepackage{amssymb}
\usepackage{amsmath}
\usepackage{tablefootnote}
\usepackage{units}
\usepackage{comment}

\usepackage{color}

\begin{document}


\title{Investigation of $^{11}$B and $^{40}$Ca levels at $8$$-$$9$~MeV by Nuclear Resonance Fluorescence}


\author{D. Gribble}
\affiliation{Department of Physics \& Astronomy, University of North Carolina at Chapel Hill, NC 27599-3255, USA}
\affiliation{Triangle Universities Nuclear Laboratory (TUNL), Duke University, Durham, North Carolina 27708, USA}

\author{C. Iliadis}
\affiliation{Department of Physics \& Astronomy, University of North Carolina at Chapel Hill, NC 27599-3255, USA}
\affiliation{Triangle Universities Nuclear Laboratory (TUNL), Duke University, Durham, North Carolina 27708, USA}

\author{R.V.F. Janssens}
\affiliation{Department of Physics \& Astronomy, University of North Carolina at Chapel Hill, NC 27599-3255, USA}
\affiliation{Triangle Universities Nuclear Laboratory (TUNL), Duke University, Durham, North Carolina 27708, USA}

\author{U. Friman-Gayer}
\affiliation{Triangle Universities Nuclear Laboratory (TUNL), Duke University, Durham, North Carolina 27708, USA}
\affiliation{Department of Physics, Duke University, Durham, North Carolina 27708-0308, USA}

\author{Krishichayan}
\affiliation{Triangle Universities Nuclear Laboratory (TUNL), Duke University, Durham, North Carolina 27708, USA}
\affiliation{Department of Physics, Duke University, Durham, North Carolina 27708-0308, USA}

\author{S. Finch}
\affiliation{Triangle Universities Nuclear Laboratory (TUNL), Duke University, Durham, North Carolina 27708, USA}
\affiliation{Department of Physics, Duke University, Durham, North Carolina 27708-0308, USA}


\date{\today}

\begin{abstract}
We report on the measurement of $^{11}$B and $^{40}$Ca levels between excitation energies of $8$ and $9$~MeV using nuclear resonance fluorescence (NRF). The experiment was carried out with nearly-monoenergetic and linearly polarized photon beams provided by the High-Intensity $\gamma$-ray Source (HI$\gamma$S) facility at the Triangle Universities Nuclear Laboratory (TUNL). States in $^{11}$B are important for calibrations of NRF measurements, while the properties of $^{40}$Ca levels impact potassium nucleosynthesis in globular clusters. For $^{40}$Ca, we report on improved excitation energies and an unambiguous $2^-$ assignment for the state at $8425$~keV. For $^{11}$B, we obtained improved values for $\gamma$-ray multipolarity mixing ratios and branching ratios of the $8920$-keV level. 
\end{abstract}


\maketitle

\section{Introduction}\label{sec:intro}
Nuclear resonance fluorescence (NRF) experiments using linearly polarized $\gamma$-ray beams allow for the measurement of level energies and widths, spins, parities, multipolarity mixing ratios, and branching ratios, and, therefore, represent an important tool for probing nuclear structure \cite{Kneissl1996,Pietralla03,zilges2022}. These properties are equally important for nuclear astrophysics, since they enter either directly in the determination of thermonuclear reaction rates, or indirectly via energy and efficiency calibrations. For example, the impact of such measurements on the rate of the $^{22}$Ne $+$ $\alpha$ s-process neutron source reactions has been demonstrated by Longland {\it et al.} \cite{Longland09}. In the latter work, only dipole excitations were observed, mostly by measuring deexcitations to the $^{26}$Mg ground state.   

The goal of the present work was primarily to investigate what else can be learned for nuclear astrophysics from NRF experiments. We report here on a measurement of $^{11}$B and $^{40}$Ca levels at energies between $8$ and $9$~MeV. Levels in the former nucleus are interesting because they often provide energy and efficiency calibrations for NRF experiments \cite{Rusev:2009eo}. States in the latter represent low-energy resonances in the $^{39}$K(p,$\gamma$)$^{40}$Ca reaction \cite{Longland18}, which is of relevance for nucleosynthesis in globular clusters \cite{Iliadis_2016,Dermigny_2017}. While the present results will likely not significantly impact the $^{39}$K $+$ $p$ reaction rate, improved information on the nuclear structure of $^{11}$B and $^{40}$Ca was obtained.

In Sec.~\ref{sec:experiment}, we describe our experimental procedure. Excitation energies and spin-parity assignments to $^{40}$Ca levels are presented in Sec.~\ref{sec:results_40ca}. Multipolarity mixing ratios of levels in $^{11}$B are discussed in Sec.~\ref{sec:results_11b}. Section~\ref{sec:summary} provides a concluding summary. In Appendix~\ref{sec:phase}, we comment on phase conventions. A reanalysis of literature data pertaining to $^{11}$B levels is given in App.~\ref{sec:app}. Our energy calibration procedure is presented in App.~\ref{sec:ener}.

\section{Experimental procedure}\label{sec:experiment}
The experiment was performed with the High-Intensity $\gamma$-ray Source (HI$\gamma$S) at the Triangle Universities Nuclear Laboratory (TUNL). Nearly-monoenergetic photon beams were produced by Compton backscattering of laser photons from relativistic electrons in a storage ring. The photon beams were linearly polarized, with the polarization vector pointing parallel to the horizontal direction. The HI$\gamma$S facility is described in detail in Ref.~\cite{Weller2009}.

The beam-defining lead collimator had a $19.05$-mm diameter and a $15.24$-cm length. Incident $\gamma$-ray beam energies used during the experiment were $8.4$, $8.6$, and $8.8$ MeV. The beam energy profile had a spread of approximately $300$ keV at full width at half maximum (FWHM). 
The beam irradiated a target, which was secured inside a plexiglass vacuum tube to reduce background counts from incident photons scattering on air. 
Two targets were used. The first consisted of boron powder ($\approx 2.1$~g), and the second of high-purity CaO powder ($\approx 6.8$~g) containing at least $99.95\%$ CaO by mass.
The target materials were placed in cylindrical polycarbonate containers, with outer dimensions of $\unit[2.2] {cm}$ diameter, $\unit[2.0] {cm}$ length, and $\unit[0.1] {cm}$ wall thickness.

\begin{figure}
\includegraphics[width=1.0\columnwidth]{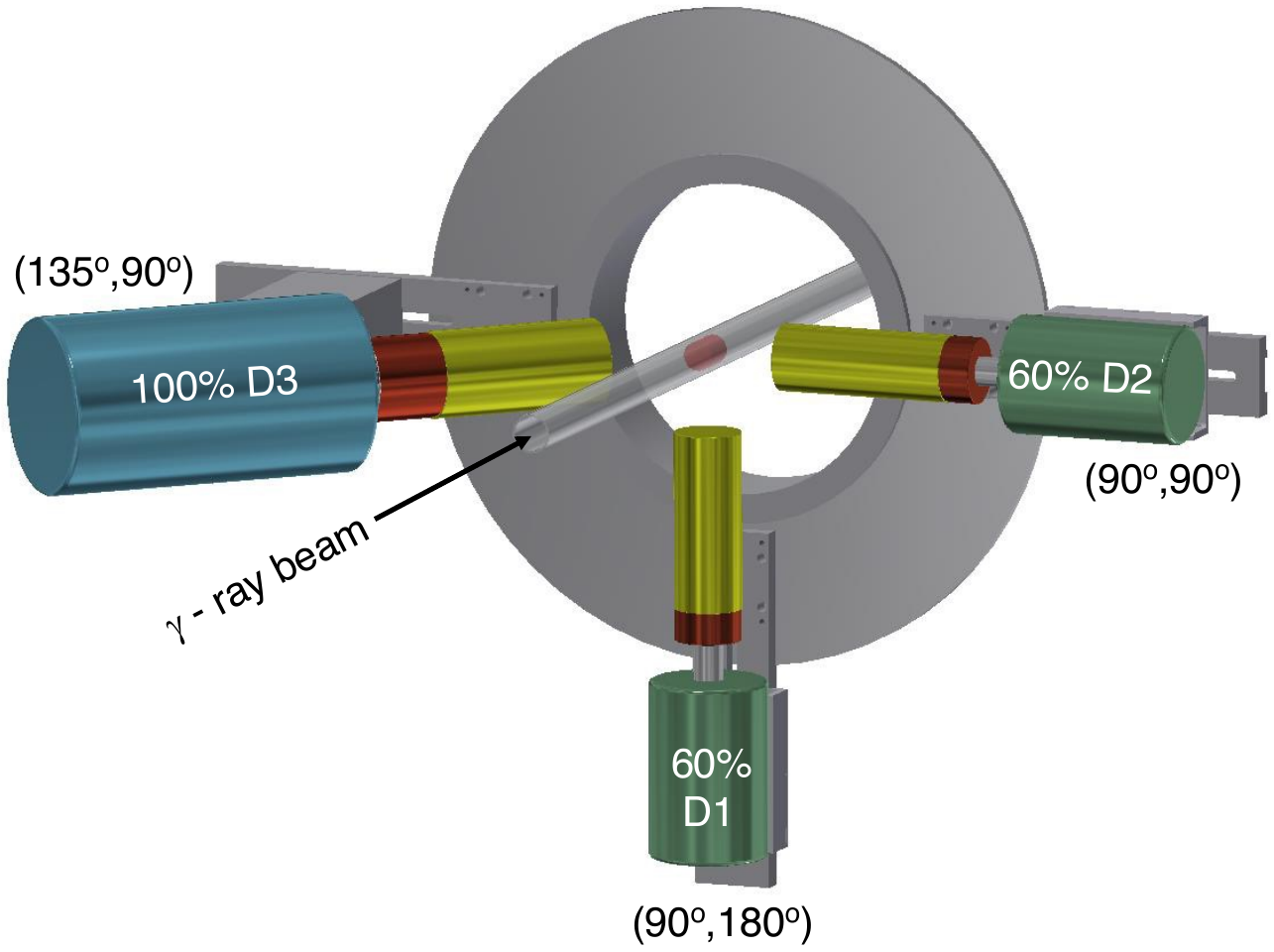}
\caption{\label{fig:setup} 
Setup used for the present experiment. The incident $\gamma$-ray beam moves inside a plexiglass vacuum tube and impinges on a sample (boron powder or CaO). The direction of the linear polarization of the beam points parallel to the horizontal plane. The dewars of the two 60\% (D1 and D2) and the 100\% (D3) HPGe detectors are colored green and blue, respectively. The front face of each detector is covered by a passive shield (yellow) to reduce backgrounds. For reference, the pairs of polar and azimuthal angles, ($\theta$, $\phi$), are depicted for each detector. The gray parts indicate the mechanical support structure of the setup.}
\end{figure}

Scattered $\gamma$ rays were detected using an array of three HPGe detectors -- two (n-type) with $60\%$ and one (p-type) with $100\%$ relative efficiency (see Fig.~\ref{fig:setup}). The two $60\%$ HPGe detectors were placed in a plane perpendicular to the beam, with a polar scattering angle of $\theta$ $=$ $90^\circ$, and azimuthal angles of $\phi$ $=$ $180^\circ$ and $90^\circ$, as measured from the vertical direction. The $100\%$ HPGe detector was arranged at $\theta$ $=$ $135^\circ$ and $\phi$ $=$ $90^\circ$. The $60\%$ and $100\%$ detectors were placed at a distance from the center of the target of $10.28$ and $15.24$~cm, respectively. All detectors were surrounded by passive shields made of copper and lead to reduce low-energy backgrounds.

For $\gamma$-ray energies below \unit[3.5]{MeV}, detector energy and efficiency calibrations were performed with  $^{60}$Co, $^{152}$Eu, and $^{56}$Co sources.  
In addition, precisely-known $\gamma$-ray energies from the decays of the $E_x$ $=$ $6792$ and $8920$-keV levels in $^{11}$B were used for energy calibration. Monte Carlo particle simulations were performed with the {\sc Geant4} toolkit \cite{Agostinelli2003, Allison2006, Allison2016} to extend the efficiency curve to higher energies. The simulated geometry included the detectors and their shielding, the plexiglass tube, and the radioactive source or target from which the gamma rays were emitted. All measured peak intensities were corrected for detector efficiency, dead time, and solid-angle attenuation effects, by using a combination of the radioactive source measurements and {\sc Geant4} simulations.

The two-step angular correlation is given by \cite{iliadis21}
\begin{align}\label{eq:angcor}
W(\theta,\phi) = A_0(1) A_0(2) + \sum\limits_{n=2,4,...}A_n(2) Q_n(k)~ \times \notag \\
\left[ A_n(1) P_n(\cos\theta) + P_{\gamma} E_n(1) P_n^{|2|}(\cos\theta) \cos(2\phi) \right] 
\end{align}
with (1) and (2) labeling the excitation and deexcitation transition, respectively; $\theta$ is the (polar) angle between the incident and emitted $\gamma$ rays; $\phi$ is the (azimuthal) angle between the vertical axis and the projection of the emitted $\gamma$-ray linear momentum onto the plane perpendicular to the incident beam direction; $A_n$ and $E_n$ are theoretical coefficients (see, e.g., Ref.~\cite{iliadis21}) that depend on the quantum numbers and multipolarity mixing ratios involved in the transition; and $P_n$ and $P_n^{|2|}$ are the unassociated and associated Legendre functions, respectively. Equation~(\ref{eq:angcor}) contains two quantities that correct for experimental artifacts: $Q_n(k)$, the solid-angle attenuation factor of detector $k$, and $P_{\gamma}$, the degree of polarization. 

The $Q$ coefficients were calculated using {\sc Geant4} by recording the initial $\gamma$-ray emission angle for each event depositing its full energy in a given detector. The attenuation coefficient of order $n$ is given by \cite{CAMP1969192}
\begin{equation}
Q_n = \frac{1}{N} \sum_{j=1}^{N} P_n(\cos\alpha_j)\label{eq:q_coefficient}
\end{equation}
with $N$ denoting the number of detected events. The initial $\gamma$-ray emission angle, $\alpha_j$, of the $j$th detected event is defined relative to a line connecting the center of the target to the center of the detector. 

The beam polarization was determined by measuring the pronounced angular correlation of the $0$~keV $\rightarrow$ $8579$~keV $(2^+)$ $\rightarrow$ $0$~keV sequence in $^{40}$Ca. We find a linear polarization of $\ge$99\%, and assumed for the subsequent analysis a value of $100$\%.

For visualizing the measured angular correlations, we introduce two quantities: the first is the analyzing power at $\theta$ $=$ $90^\circ$, defined by \cite{iliadis21}
\begin{equation}
A(\theta = 90^\circ) = \frac{W(90^\circ, 180^\circ) - W(90^\circ, 90^\circ)}{W(90^\circ, 180^\circ) + W(90^\circ, 90^\circ)} \label{eq:analyzing}
\end{equation}
and the second is the ratio, $R$, involving the detector at ($\theta$, $\phi$) $=$ (135$^\circ$, 90$^\circ$),
\begin{equation}
R = \frac{W(135^\circ, 90^\circ)}{W(90^\circ, 180^\circ) + W(90^\circ, 90^\circ)} \label{eq:analyzing2}.
\end{equation}
Below, we will display our data, together with theoretical predictions, in the $R$ {\it vs.} $A$ plane.

For the extraction of multipolarity mixing ratios, previous works have frequently applied an approximate analysis procedure (see, e.g., Ref.~\cite{Rusev:2009eo}). With this ``min-max'' method, each calculation is performed three times, i.e., for the mean value of the measured analyzing power and for both of its experimental bounds. The range of mixing-ratio values is then obtained from the intersections of the measured analyzing power (1$\sigma$ region) with the theoretically predicted curve. This procedure is not rigorous from a statistical point of view, and can lead to unreliable results, especially if the 1$\sigma$ uncertainties of the measured analyzing powers are significant. In this work, we will instead analyze the data using a rigorous method based on Bayesian statistics, and estimate mixing ratio values from posterior probability densities \cite{2017bmad}.

\section{Results for $^{40}$Ca}\label{sec:results_40ca}
\subsection{Excitation energies}
One goal of the present work was to derive precise excitation energies of $^{40}$Ca levels that may contribute to the $^{39}$K(p,$\gamma$)$^{40}$Ca thermonuclear reaction rate. Consider, as an example, the $96$-keV resonance (center-of-mass energy), corresponding to the $8425$-keV level (see Tab.~\ref{tab:40ca_energies}), which dominates the total reaction rates in the temperature region of $20$ $-$ $80$~MK (see Fig. 5 in Ref.~\cite{Longland18}). If the energy of this resonance is varied by, say, $0.5$~keV (or $1.0$~keV), and all other parameters are kept at their same values, the total rate at $40$~MK would change by $16$\% (or $34$\%). This sensitivity results from the fact that the energies enter exponentially in the expression of the narrow-resonance reaction rates \cite{Iliadis_2015}, and emphasizes the importance of a reliable energy calibration. Our calibration procedure is described in detail in App.~\ref{sec:ener}. Results for $^{40}$Ca level energies are listed in Tab.~\ref{tab:40ca_energies} and compared to previous values provided by the Evaluated Nuclear Structure Data File (ENSDF) \cite{ensdf2021_40}. 

Present (column 3) and previous (column 4) measured $\gamma$-ray energies agree within 1$\sigma$ only for the $8092$~keV $(2^+)$ $\rightarrow$ $0$~keV transition, although our uncertainty is significantly larger. For the $8425$~keV $(2^-)$ $\rightarrow$ $3737$~keV transition, no uncertainty is provided in Ref.~\cite{ensdf2021_40} in its table of ``adopted Levels, Gammas'' (however, see footnote ``g'' in Tab.~\ref{tab:40ca_energies}). For the remaining three transitions, $8579$~keV $(2^+)$ $\rightarrow$ $0$~keV, $8748$~keV $(2^+)$ $\rightarrow$ $0$~keV, and $8982$~keV $(2^+)$ $\rightarrow$ $0$~keV, our $\gamma$-ray energies have smaller uncertainties compared to the evaluation of Ref.~\cite{ensdf2021_40}.

\begin{figure}
\includegraphics[width=1.0\columnwidth]{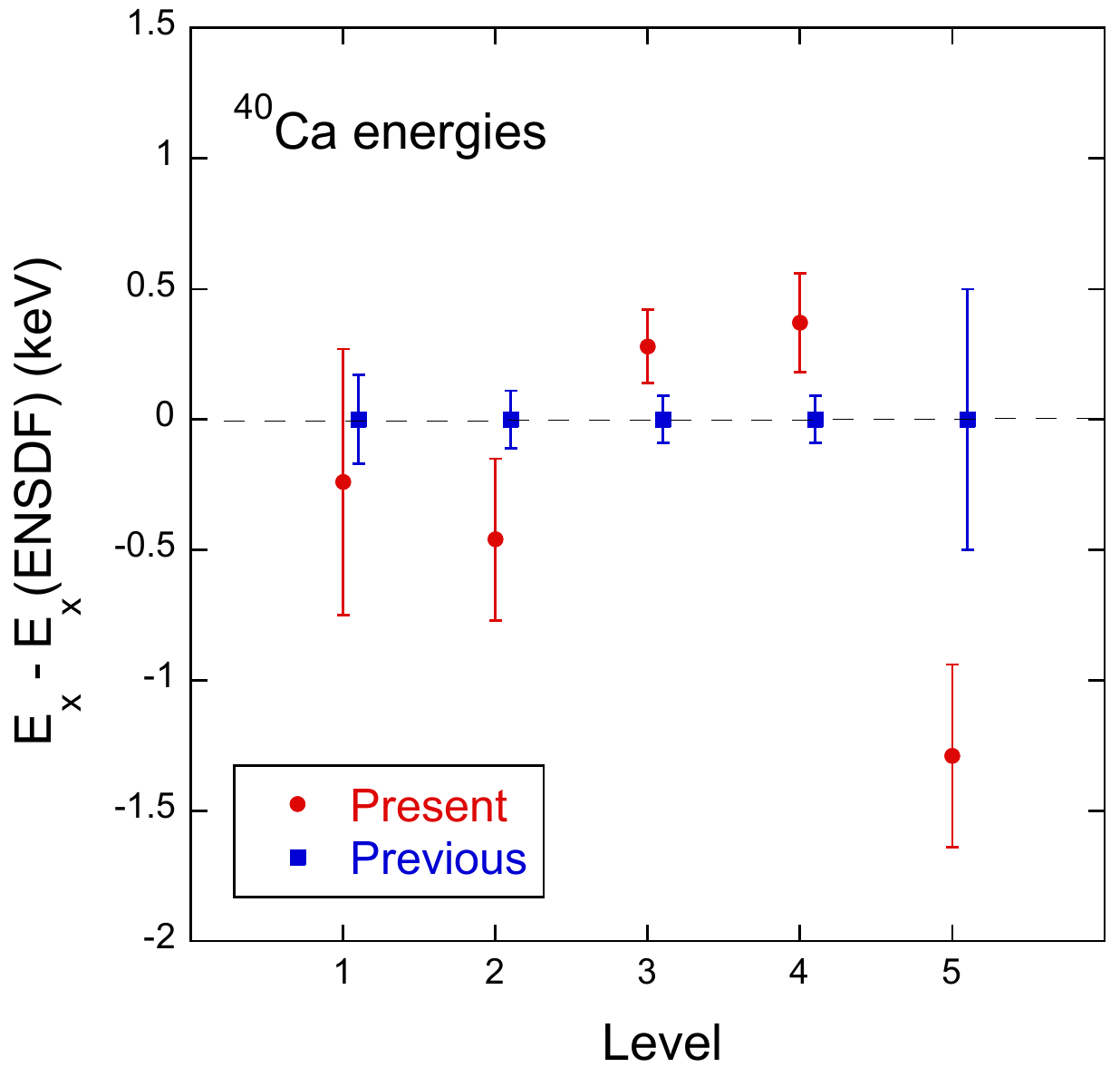}
\caption{\label{fig:40caener} 
Excitation energies of $^{40}$Ca levels observed in the present work (red circles) as compared to the previously evaluated results \cite{ensdf2021_40} (blue squares). For better comparison, the ordinate refers to the difference relative to the previous mean value. The state labels on the abscissa correspond to those listed in column 1 of Tab.~\ref{tab:40ca_energies}.
}
\end{figure}

We derived $^{40}$Ca excitation energies by applying the recoil correction to the measured $\gamma$-ray energies. No corrections for Doppler shifts were needed because the detector was located at an angle of $\theta$ $=$ $90^\circ$. The results are listed in column 5 of Tab.~\ref{tab:40ca_energies}. Column 6 provides the literature values, which represent the evaluated and recommended values based on a number of previous experiments. The differences, $E_x^{\text{present}}$ $-$ $E_x^{\text{ENSDF}}$, are displayed in Fig.~\ref{fig:40caener}. As was the case for the $\gamma$-ray energies, the present and previous excitation energies agree within 1$\sigma$ only for the $8092$~keV $(2^+)$ level (label ``1''). For the $8982$~keV $(2^+)$ state (label ``5''), the mean values differ by $1.3$~keV and our uncertainty ($\pm 0.35$~keV) is smaller than the previously evaluated result ($\pm 0.5$~keV).

\begin{table*}[]
\begin{center}
\caption{$^{40}$Ca excitation energies.}\label{tab:40ca_energies}
\begin{ruledtabular}
\begin{tabular}{l l l l l c}
Level\footnotemark[1]   &   Transition (keV)  &   $E_{\gamma}^{\text{present}}$ (keV)\footnotemark[2] &      $E_{\gamma}^{\text{ENSDF}}$ (keV)\footnotemark[6]    &   $E_x^{\text{present}}$ (keV)\footnotemark[5]   &    $E_x^{\text{ENSDF}}$ (keV)\footnotemark[6]  \\
\hline
1   &   $8092$ $(2^+)$ $\rightarrow$ $0$    &   8090.49$\pm$0.51                  &  8090.6$\pm$0.2 &   8091.37$\pm$0.51    &   8091.61$\pm$0.17  \\
2   &   $8425$ $(2^-)$ $\rightarrow$ $3737$ &   4687.37$\pm$0.31                  &  4687.8\footnotemark[7]         &   8424.35$\pm$0.31\footnotemark[8]    &   8424.81$\pm$0.11  \\
3   &   $8579$ $(2^+)$ $\rightarrow$ $0$    &   8578.10$\pm$0.14\footnotemark[3]  &  8577.7$\pm$0.2 &   8579.08$\pm$0.14    &   8578.80$\pm$0.09  \\
4   &   $8748$ $(2^+)$ $\rightarrow$ $0$    &   8747.55$\pm$0.19\footnotemark[4]  &  8748.4$\pm$0.2 &   8748.59$\pm$0.19    &   8748.22$\pm$0.09  \\
5   &   $8982$ $(2^+)$ $\rightarrow$ $0$    &   8980.13$\pm$0.35                  &  8981.4$\pm$0.5 &   8981.21$\pm$0.35    &   8982.5$\pm$0.5  \\
\end{tabular}
\end{ruledtabular}
\footnotetext[1]{Level numbers correspond to those shown in Fig.~\ref{fig:40caener}.} 
\footnotetext[2]{Values do not include the recoil correction; no correction for the Doppler shift is required (see text).} 
\footnotetext[3]{Weighted average of independent measurements performed at three different incident beam energies.} 
\footnotetext[4]{Weighted average of independent measurements performed at two different incident beam energies.} 
\footnotetext[5]{Present results; uncertainties include those for peak centroids and the energy calibration (common uncertainty: $\pm 0.10$~keV).} 
\footnotetext[6]{According to Ref.~\cite{ensdf2021_40}, ``...values with uncertainties are averaged values from different $\gamma$-ray studies. A large number of values without uncertainties are from $^{39}$K(p,$\gamma$), which are from level-energy differences since most $\gamma$-ray energies are not available.''}
\footnotetext[7]{Ref.~\cite{ensdf2021_40} lists $4688.2\pm1.5$~keV from (p,p$^{\prime}\gamma$) in the comment section of their table.}
\footnotetext[8]{Calculated from the $\gamma$-ray energy in column 3, the recoil correction, and the energy of the final excited state, $E_x$ $=$ $3736.69\pm0.05$~keV \cite{ensdf2021_40}.}
\end{center}
\end{table*}

\subsection{Spin-parity of the $8425$-keV state}
The $8425$-keV level in $^{40}$Ca corresponds to a $p$-wave resonance in the $^{39}$K(p,$\gamma$)$^{40}$Ca reaction at a center-of-mass energy of $96$~keV. The spin-parity is listed as $J^\pi$ $=$ $2^-$ in Ref.~\cite{ensdf2021_40}, and the level decays mainly to the $3737$-keV ($3^-$) state. The $8425$-keV level dominates the thermonuclear reaction rates below a stellar temperature of $100$~MK (see Fig.~5 in Ref.~\cite{Longland09}). If this level had $J^\pi$ $=$ $2^+$ instead, it would correspond to an $s$-wave resonance, implying a dominant $^{39}$K(p,$\gamma$)$^{40}$Ca rate contribution at even higher temperatures. In addition, a $2^+$ (natural parity) assignment would imply a contribution to the $^{39}$K(p,$\alpha$)$^{36}$Ar rate. Since this spin-parity assignment has significant implications for potassium nucleosynthesis in globular cluster stars \cite{Iliadis_2016}, it must be considered carefully.

Arguments for a $2^-$ assignment to the $8425$-keV level are presented by ENSDF \cite{ensdf2021_40}. They are based on measurements of the $^{41}$Ca($^3$He,$\alpha$)$^{40}$Ca pick-up reaction \cite{CLINE197491}, $^{39}$K(d,n)$^{40}$Ca \cite{FUCHS1969545} and $^{39}$K($^3$He,d)$^{40}$Ca \cite{Erskine66,Seth67} transfer reactions, inelastic scattering of polarized protons \cite{Hosono82,Horen85}, and inelastic electron scattering \cite{STEFFEN198023}. However, neither the $\ell$ $=$ $2$ transfer observed in Ref.~\cite{CLINE197491}, nor the mixed transfer of $\ell$ $=$ $1$ $+$ $3$ observed in Ref.~\cite{FUCHS1969545} result unambiguously in a $2^-$ assignment. Erskine \cite{Erskine66} reports a preliminary assignment of ``($2^-$),'' but locates the level at $8465\pm12$~keV, about $40$~keV higher than the values listed in Tab.~\ref{tab:40ca_energies}. He states that, because of many contaminant peaks in this region of the spectrum, ``identification of the (...) $8.465$-MeV levels as states in $^{40}$Ca is a little uncertain.'' Seth {\it et al.} \cite{Seth67} reported a $2^-$ assignment, but the fact that their angular distribution data were consistent with an $\ell$ $=$ $3$ fit does not unambiguously determine the spin-parity. Furthermore, they write ``Unfortunately, the shape of the measured angular distribution is not very well determined.'' The $^{40}$Ca($\vec{p},p^\prime)^{40}$Ca study by Hosono {\it et al.} \cite{Hosono82} did not determine a $2^-$ assignment, but rather assumed this value, based on an earlier compilation \cite{EVL1973}. In addition, their analyzing power fit does not describe the data well. Horen {\it et al.} \cite{Horen85} reported much improved $^{40}$Ca($\vec{p},p^\prime)^{40}$Ca data consistent with a $2^-$ assignment, but their overall energy resolution was only $80$~keV at FWHM. Similarly, the $^{40}$Ca($e$,$e^\prime$)$^{40}$Ca study of Ref.~\cite{STEFFEN198023} measured data consistent with $2^-$, but does not provide proof for an unambiguous assignment.  

Considering all the evidence together, it is likely that the $8425$-keV state has a spin-parity of $2^-$. This conclusion is consistent with the shell-model interpretation of this state being the $J$ $=$ $2$, $T$ $=$ $1$ level of the $(d_{3/2}^{-1} f_{7/2})$ configuration, since the observed excitation energy agrees with those of the $2^-$ isobaric analog states in $^{40}$K and $^{40}$Sc. However, it is worthwhile to determine the spin-parity unambiguously from a single measurement.

\begin{figure*}
\includegraphics[width=2.0\columnwidth]{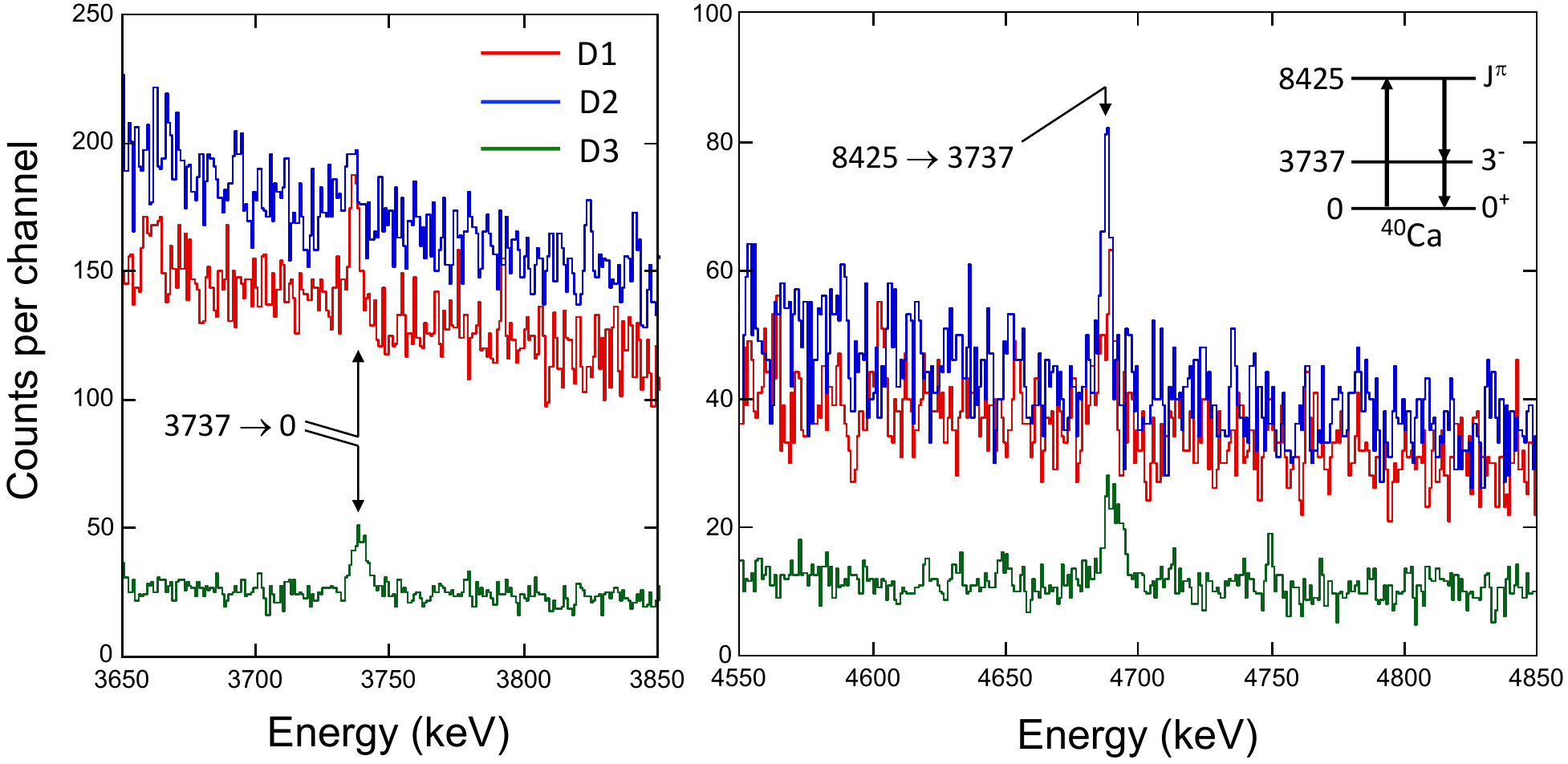}
\caption{\label{fig:spec40ca} 
Pulse height spectra measured at an incident $\gamma$-ray energy of $8400$~keV. The spectra of the vertical (D1), horizontal (D2), and out-of-plane (D3) HPGe detectors are depicted in red, blue, and green, respectively. See also Fig.~\ref{fig:setup}. The peaks are labeled according to the initial and final state energies (given in units of keV) of the transition. The spectrum in green was scaled down by a factor of $\approx$2 to account for the ratio of detection efficiencies; detectors D1 and D2 had similar efficiencies. (Left panel) Energy region of the secondary decay from the $E_x$ $=$ $3737$~keV level in $^{40}$Ca to the ground state, $3737$~keV $\rightarrow$ $0$~keV. (Right panel) Energy region of the primary transition from the $E_x$ $=$ $8425$~keV level to the $E_x$ $=$ $3737$~keV state, $8425$~keV $\rightarrow$ $3737$~keV.
}
\end{figure*}

Relevant parts of our pulse-height spectra, measured at an incident $\gamma$-ray energy of $8400$~keV, are presented in Fig.~\ref{fig:spec40ca}. The spectra of the vertical (D1), horizontal (D2), and out-of-plane (D3) HPGe detectors are depicted in red, blue, and green, respectively. The peaks in the spectra correspond to the transition from the $E_x$ $=$ $8425$~keV level in $^{40}$Ca to the state at $E_x$ $=$ $3737$~keV, and the subsequent decay to the ground state. The unambiguous identification of the two peaks with the excitation of the $E_x$ $=$ $8425$-keV level in $^{40}$Ca is based on the following arguments. First, the measured $\gamma$-ray energies agree with those of the expected primary ($8425$~keV $\rightarrow$ $3737$~keV) and secondary ($3737$~keV $\rightarrow$ $0$~keV) transitions. Second, we observed the two peaks only at an incident energy of $8400$~keV; i.e., at the higher bombarding energies of $8600$~keV or $8800$~keV these peaks are absent. (Recall that our incident beam resolution was about $300$~keV; see Sec.~\ref{sec:experiment}.) Third, the only other $^{40}$Ca states that could have been populated at $8400$~keV bombarding energy, and potentially fed either the $E_x$ $=$ $8425$-keV level or the $3737$-keV state, are located at $E_x$ $=$ $8484$~keV (1$^-$, 2$^-$, 3$^-$), $8540$~keV (1, 2$^+$), and $8579$~keV (2$^+$) \cite{ensdf2021_40}:

(i) Only a ground state transition has been reported for $E_x$ $=$ $8579$~keV, and we observed this decay at $8400$~keV bombarding energy. If this level would feed the $E_x$ $=$ $8425$~keV or $3737$~keV states at $8400$~keV bombarding energy via so far unobserved branches, we should have observed the decays of the latter two states also at $8600$~keV bombarding energy, where the $8579$-keV level is more strongly populated. However, as already mentioned above, no such decays have been observed at this higher bombarding energy;

(ii) Similarly, the $E_x$ $=$ $8540$~keV level has known branches to the ground state and to the state at $3353$~keV. We observed none of these transitions. Again, if the $8540$-keV level would feed the $E_x$ $=$ $8425$-keV or $3737$-keV states at $8400$~keV bombarding energy via so far unobserved branches, we should have observed the decays of the latter two states also at $8600$~keV bombarding energy, but we did not;  

(iii) The $E_x$ $=$ $8484$~keV level has reported branches to the $3737$- and $5903$-keV states. However, we did not observe the corresponding primary decays feeding these levels (see Fig.~\ref{fig:spec40ca}). A significant, and so far unobserved, transition from  the $8484$-keV state to the $8425$-keV level would imply a transition energy of only $59$~keV. We find this highly unlikely, considering that the decay of the $8484$-keV state can proceed, in principle, via $E1$, $M1$, or $E2$ multipolarities to about $40$ lower-lying levels with much larger transition energies. Furthermore, the energy of the $8484$-keV level is about half-way between the $8400$ and $8600$~keV bombarding energies. If this level did indeed feed the $8425$-keV state at $8400$~keV bombarding energy, such feeding would also have occurred at $8600$~keV bombarding energy. Again, we observed the $8425$~keV $\rightarrow$ $3737$~keV and $3737$~keV $\rightarrow$ $0$~keV transitions only at $8400$~keV bombarding energy.

We did not detect any of the other known decays of the $E_x$ $=$ $8425$~keV level ($8425$~keV $\rightarrow$ $5903$~keV and $8425$~keV $\rightarrow$ $6025$~keV), since they have much smaller reported branching ratios ($24$\% and $19$\%, respectively \cite{ensdf2021_40}). Their primary-transition energies are smaller, resulting in a higher detection efficiency, but also a higher background, compared to the energy region of the observed transitions.

\begin{figure}
\includegraphics[width=1.0\columnwidth]{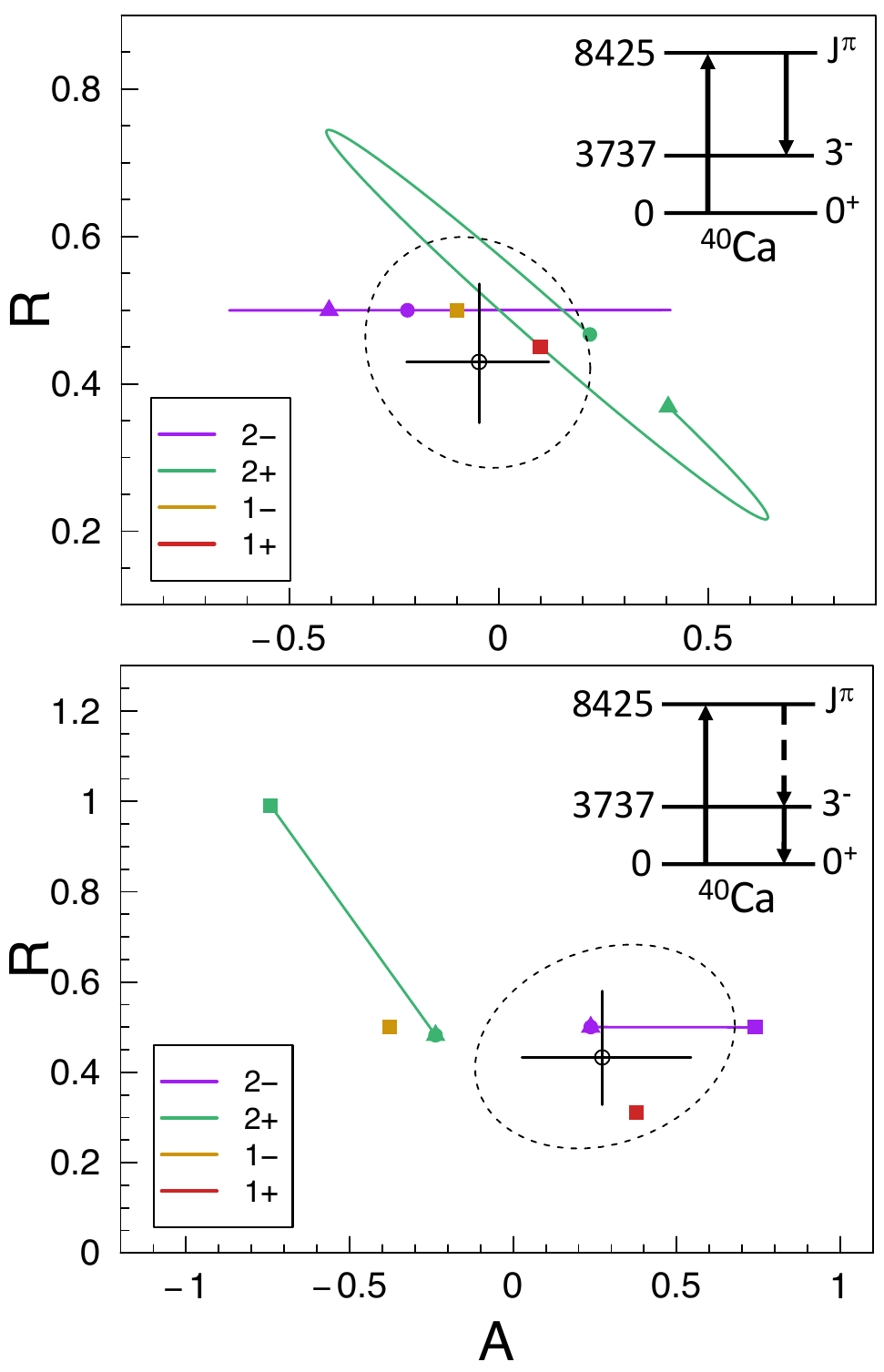}
\caption{\label{fig:RvsA40ca} 
Ratio {\it vs.} analyzing power for the deexcitation of the $8425$-keV level in $^{40}$Ca. (Top) Primary transition, $8425$~keV $\rightarrow$ $3737$~keV. (Bottom) Secondary transition, $3737$~keV $\rightarrow$ $0$~keV, with the primary transition unobserved. The measured values are shown in black, where both the error bars and the dashed black line (error ellipse) correspond to 68\% coverage probability. Different colors refer to theoretical values based on given choices of $J^\pi$ for this level. The purple and green lines correspond to the theoretical results for a range of multipolarity mixing ratios of the measured transitions. Colored circles, squares, and triangles indicate theoretical values for mixing ratios of $\delta$ $=$ $-10$, $0$, and $+10$, respectively. For $J^\pi$ $=$ $1^-$ and $1^+$, only the values corresponding to $\delta$ $=$ $0$ are indicated (see text). Notice that, in the top panel, the purple and green squares (signifying $\delta$ $=$ $0$) are covered by the yellow and red ones, respectively.
}
\end{figure}

Based on our measured intensities, Fig.~\ref{fig:RvsA40ca} presents the ratio, $R$, versus the analyzing power, $A$, for the $8425$~keV $\rightarrow$ $3737$~keV primary transition (top) and the $3737$~keV $\rightarrow$ $0$~keV secondary one with the primary transition unobserved (bottom). The measured data are given in black, with the dashed line depicting the associated 68\% error ellipse. Colors signify theoretical predictions based on different choices of $J^\pi$ values for the $8425$-keV level. Only predictions for dipole and quadrupole transitions (i.e., $J^\pi$ $=$ $1^\pm$, $2^\pm$ ) are shown, as higher multipoles are unlikely. The colored circles, squares, and triangles depict theoretical predictions for mixing ratios of $\delta$ $=$ $-10$, $0$, and $+10$, respectively. For $J^\pi$ $=$ $1^\pm$, we only present the results for pure transitions ($\delta$ $=$ $0$), as mixed decays would imply multipoles higher than quadrupole radiation.

The measurement of the primary transition (top panel), for which we found values of $A$ $=$ $-0.05\pm0.18$ and $R$ $=$ $0.43^{+0.11}_{-0.08}$, does not allow for an unambiguous $J^\pi$ assignment because the predictions for all spin-parity choices, $1^+$, $1^-$, $2^+$, or $2^-$, intersect with the experimental error ellipse. However, the measurement of the secondary transition (bottom panel), for which we obtained $A$ $=$ $+0.27\pm0.27$ and $R$ $=$ $0.43^{+0.15}_{-0.10}$, excludes both the $J^\pi$ $=$ $2^+$ and $1^-$ values. We can also exclude $1^+$, as it would imply an $M2$ transition to the $3737$-keV level. Since the $8424$-keV state can, in principle, decay to many lower lying $^{40}$Ca levels via $E1$, $M1$, or $E2$ transitions, an $M2$ transition for the strongest $\gamma$-ray branch is highly unlikely. This leaves $J^\pi$ $=$ $2^-$ as the only possibility consistent with our data, in agreement with the result evaluated in Ref.~\cite{ensdf2021_40}. Although the $3737$-keV level is fed by the side-branch $8425$~keV $\rightarrow$ $6025$~keV $\rightarrow$ $3737$~keV, its branching ratio is small ($10$\%) and its effect is contained in the error bars of the measured peak intensities.

We close this section by noting that populating the $8424$-keV state in our NRF experiment is not fundamentally surprising. For the corresponding $M2$ ground-state transition, Fagg et al. \cite{Fagg1971} measured a $\gamma$-ray width of $2.6^{+10}_{-8} \times 10^{-2}$~eV in their $(e,e^\prime)$ experiment\footnote{This experimental ground-state width was not adopted by ENSDF \cite{ensdf2021_40} and the reason can be traced back to Ref.~\cite{ENDT19781}, who wrote in the footnote to their Tab.~40.17, ``The ground-state width of $2.6 \pm 0.9$~eV in [Fagg et al.] from $(e,e^\prime)$, leading to an $M2$ strength of $350$~W.u., should be rejected as erroneous." Unfortunately, Ref.~\cite{ENDT19781} adopted a value that is erroneous by two orders of magnitude.}. This value corresponds to $3.5$~W.u., which is smaller than the recommended upper limit (RUL) of $5$~W.u. for $M2$ isospin-allowed ($\Delta T$ $=$ $1$) transitions \cite{ENDT1993171}.

\section{Results for $^{11}$B}\label{sec:results_11b}
The $^{11}$B ground-state spin-parity is $3/2^-$ and, unlike the case of $^{40}$Ca, transitions can proceed via multipolarity mixing in both the first (excitation) and second (deexcitation) transition. The utility of the $^{11}$B($\gamma$,$\gamma$)$^{11}$B reaction as a calibration standard and, in particular, the reliable knowledge of the mixing ratios for measurements of the photon flux in NRF experiments has been pointed out by Rusev {\it et al.} \cite{Rusev:2009eo}. The latter work was performed at the HI$\gamma$S facility with a detection setup similar to ours, except that all the detectors were positioned in a plane perpendicular to the beam ($\theta$ $=$ $90^\circ$). In other words, only the analyzing power, $A$, was measured, but not the ratio, $R$ (see Eqs.~(\ref{eq:analyzing}) and (\ref{eq:analyzing2})). We were particularly interested in mixed transitions from the well-known state at $E_x$ $=$ $8920.47\pm0.11$~keV ($5/2^-$) \cite{ensdf2021_11}. This level decays mainly to the ground state, but also has a small branch to the $4445$-keV ($5/2^-$) state. Since the ground-state spin of $^{11}$B is $3/2$, both the excitation and deexcitation transitions can be of mixed multipolarity, requiring two mixing ratios in Eq.~(\ref{eq:angcor}). 

\begin{figure*}
\includegraphics[width=2.0\columnwidth]{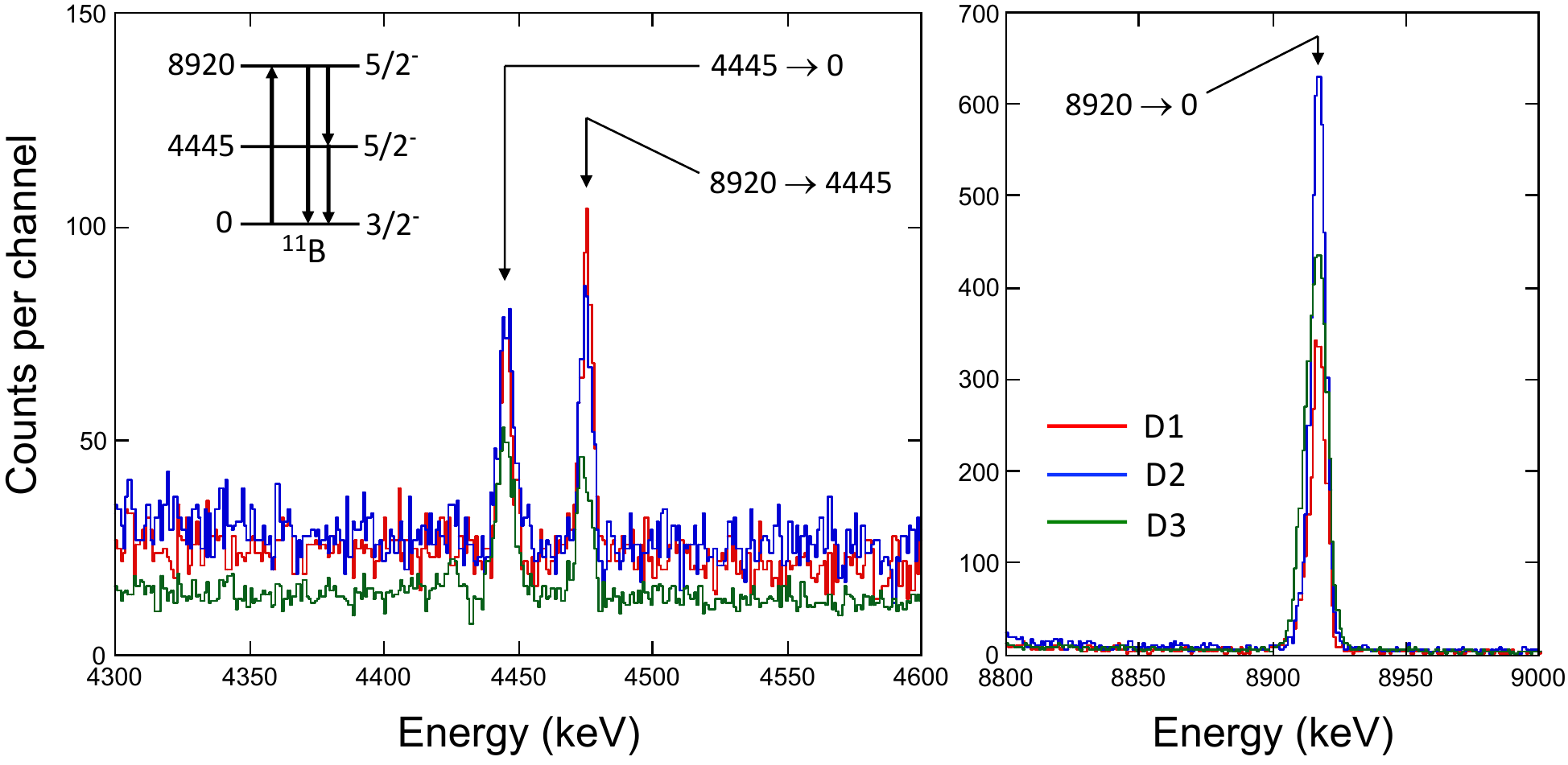}
\caption{\label{fig:b11spec} 
Pulse height spectra measured at an incident $\gamma$-ray energy of $8920$~keV. The spectra of the vertical (D1), horizontal (D2), and out-of-plane (D3) HPGe detectors are depicted in red, blue, and green, respectively. See also Fig.~\ref{fig:setup}. The peaks are labeled according to the initial and final state energies (given in units of keV) of the transition. The spectrum shown in green was scaled down by a factor of $\approx$2 to account for the ratio of detection efficiencies; detectors D1 and D2 had similar efficiencies. (Left panel) Energy region of the primary transition to the $E_x$ $=$ $4445$~keV level in $^{11}$B, $8920$~keV $\rightarrow$ $4445$~keV, and the subsequent secondary transition, $4445$~keV $\rightarrow$ $0$~keV. (Right panel) Energy region of the primary transition to the $^{11}$B ground state, $8920$~keV $\rightarrow$ $0$~keV.} 
\end{figure*}

Relevant parts of our pulse-height spectra, measured at an incident $\gamma$-ray energy of $8920$~keV, are presented in Fig.~\ref{fig:b11spec}. The spectra of the vertical (D1), horizontal (D2), and out-of-plane (D3) HPGe detectors are depicted in red, blue, and green, respectively. The left panel represents the energy region of the $8920$~keV $\rightarrow$ $4445$~keV and $4445$~keV $\rightarrow$ $0$~keV transitions, whereas the right panel displays the energy region of the $8920$~keV $\rightarrow$ $0$~keV transition. 

Unfortunately, the results published in Ref.~\cite{Rusev:2009eo} are subject to an inconsistent phase definition in the angular correlation expressions they adopted, as explained in more detail in App.~\ref{sec:phase}. The erroneous mixing ratio values have also been adopted in ENSDF \cite{ensdf2021_11}. Therefore, we reanalyzed the data of Ref.~\cite{Rusev:2009eo} and extracted from their published asymmetries (column 3 of their Tab.~I) the allowed mixing-ratio ranges. The results are presented in  App.~\ref{sec:app}. We will compare below our results with these ``corrected Rusev {\it et al.} values.''

Present and previous results for the $\gamma$-ray multipolarity mixing ratios and branching ratios are listed in Tab.~\ref{tab:11b_delta}.

\subsection{Transition of the $8920$-keV level to the ground state}\label{sec:8920to0}
For the $8920$-keV ground-state transition, Comsan {\it et al.} \cite{Comsan68} reported an $E2/M1$ multipolarity mixing ratio of $\delta^{\text{Comsan}}$ $=$ $-0.11\pm0.04$, which was derived from their $^{10}$B(d,p$\gamma$)$^{11}$B experiment. Unfortunately, they did not state their adopted phase convention for the definition of $\delta$. Rusev {\it et al.} \cite{Rusev:2009eo} reported a value of $\delta^{\text{Rusev}}$ $=$ $0.000\pm0.014$. However, without additional assumptions, their measurement restricted the mixing ratio to two ranges, one near $\delta^{\text{Rusev}}$ $=$ $-1.3$ and one near $0$. They excluded the former solution ``because [it] leads to $E2$ admixtures that are unusually large compared with known values'' \cite{Rusev:2009eo}. As already pointed out above, their reported numerical results are erroneous. The reanalysis of their data (see App.~\ref{sec:app} and Tab.~\ref{tab:app}) yields four, instead of two, possible solutions: $\delta^{\text{Rusev}} = $ $-6.7^{+0.9}_{-1.8}$, $-0.387^{+0.019}_{-0.018}$, $-0.001^{+0.027}_{-0.028}$, and $+3.06^{+0.28}_{-0.19}$. While it may be reasonable to exclude the first and last solutions because of implied unusually large $E2$ admixtures, disregarding the second solution without further evidence is questionable.

\begin{figure}
\includegraphics[width=1.0\columnwidth]{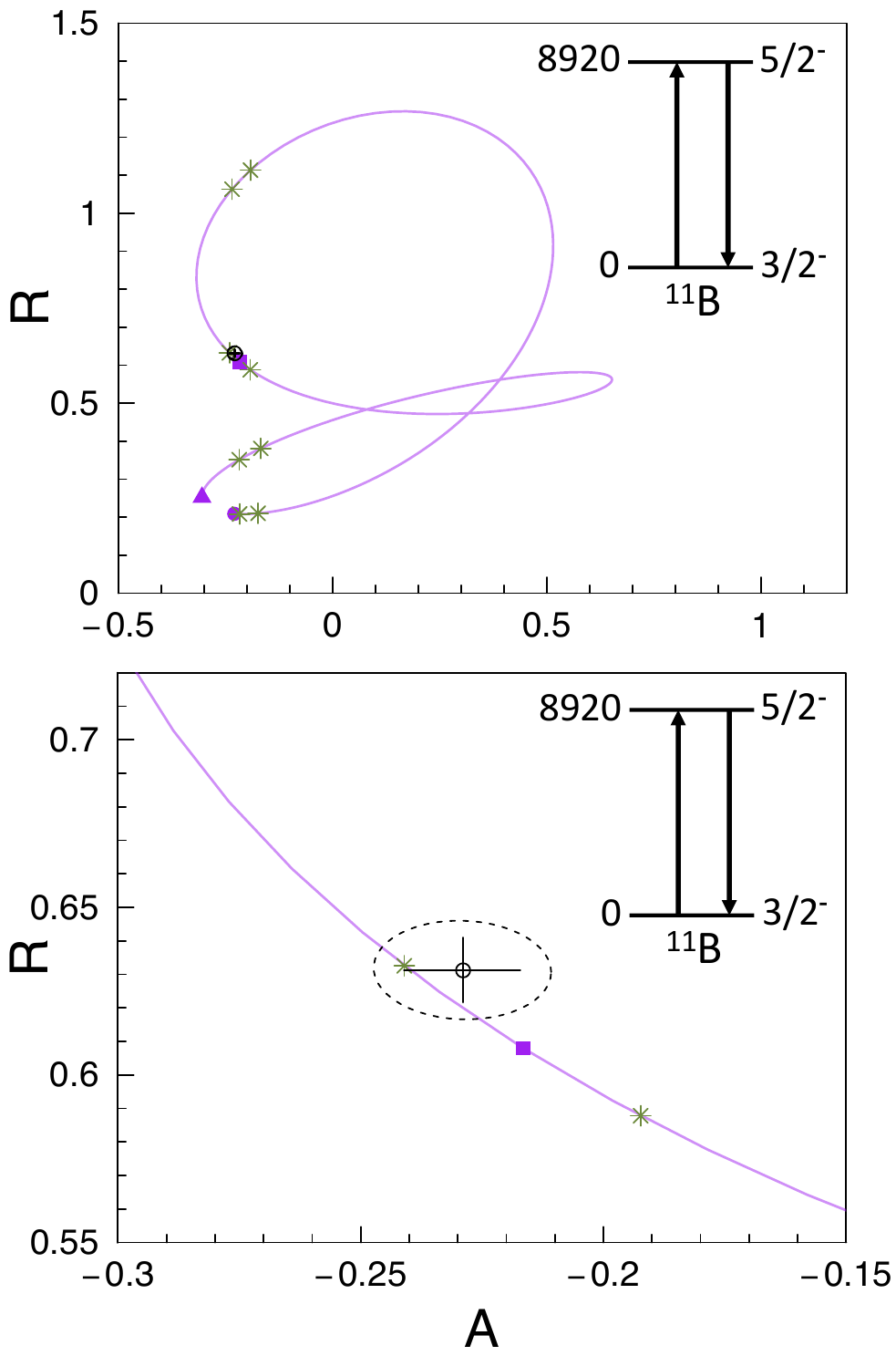}
\caption{\label{fig:b11AC} 
Ratio {\it vs.} analyzing power for the deexcitation of the $8920$-keV ($5/2^-$) level to the $^{11}$B ground state ($3/2^-$). (Top) The measured value is depicted in black. The purple line depicts the theoretical solutions for the full range of mixing ratios. Purple circles, squares, and triangles indicate theoretical values of $\delta$ $=$ $-10$, $0$, and $+10$, respectively. The regions between pairs of green asterisks correspond to the four possible solutions from our reanalysis of the data by Ref.~\cite{Rusev:2009eo} (see App.~\ref{sec:app}). (Bottom) Expanded view of the top panel. Both the error bars and the dashed black line (error ellipse) refer to 68\% coverage probability.
}
\end{figure}

Our experimental results are presented in Fig.~\ref{fig:b11AC}. The purple line depicts the theoretical solutions for a range of mixing ratios, $\delta$. Purple circles, squares, and triangles indicate theoretical values of $\delta$ $=$ $-10$, $0$, and $+10$, respectively. The four regions marked by pairs of green asterisks correspond to the possible solutions based on the reanalysis of the Rusev {\it et al.} \cite{Rusev:2009eo} data (see App.~\ref{sec:app} and Tab.~\ref{tab:app}). The result from the present experiment is presented as a black circle. It can be seen that our data ($A$ $=$ $-0.23\pm0.01$ and $R$ $=$ $0.63\pm0.01$) provide a unique solution near a value of zero (purple square), without requiring any additional assumptions. In particular, we can exclude three of the solutions resulting from the data of Ref.~\cite{Rusev:2009eo}. 

The bottom panel presents an expanded view of the top one. It is apparent that our measurement gives a significantly more precise value for the mixing ratio compared to the solution near zero resulting from the data of Ref.~\cite{Rusev:2009eo}. By applying a Bayesian analysis directly to the measured intensities, we obtained a value of $\delta^{\text{present}}$ $=$ $-0.023 \pm 0.009$ (68\% coverage probability; see Tab.~\ref{tab:11b_delta}). All signs of mixing ratios quoted in the present work are consistent with the Steffen-Krane phase convention (see App.~\ref{sec:phase} for more details on phase conventions). Consequently, our data predict a $\approx$0.05\% $E2$ contribution to the total transition strength. 

\begin{table*}[]
\begin{center}
\caption{Multipolarity mixing ratios (consistent with the Steffen-Krane phase convention, unless mentioned otherwise) and branching ratios for decaying levels in $^{11}$B.} \label{tab:11b_delta}
\begin{ruledtabular}
\begin{tabular}{l c c c c}
   Transition~(keV)  &    $\delta^{\text{present}}$\footnotemark[1] &         $\delta^{\text{lit}}$\footnotemark[2]  & $B_\gamma^{\text{present}}$~(\%) &     $B_\gamma^{\text{ENSDF}}$~(\%)\footnotemark[5]    \\
\hline
$8920$ ($5/2^-$) $\rightarrow$ $0$ ($3/2^-$)    &     $-0.023 \pm 0.009$    &   $-0.11 \pm 0.04$\footnotemark[3]        &  $96.83\pm0.13$   &      $95.0\pm1.0$    \\
$8920$ ($5/2^-$) $\rightarrow$ $4445$ ($5/2^-$) &     $+0.72 \pm 0.40$      &                                           &   $3.17\pm0.13$   &      $4.5\pm0.5$      \\
$4445$ ($5/2^-$) $\rightarrow$ $0$ ($3/2^-$)    &                           &     $+0.19\pm0.03$\footnotemark[4]        &   $100$           &      $100$           \\
\end{tabular}
\end{ruledtabular}
\footnotetext[1]{From present work. Uncertainties correspond to coverage probabilities of 68\%.} 
\footnotetext[2]{The mixing ratio values published in Ref.~\cite{Rusev:2009eo} are erroneous (see App.~\ref{sec:phase}). A reanalysis of their data is presented in  App.~\ref{sec:app}.} 
\footnotetext[3]{From Comsan {\it et al.} \cite{Comsan68}; the phase convention is not mentioned explicitly.} 
\footnotetext[4]{From Bell {\it et al.} \cite{BELL1968481}. Their published value has the opposite sign because they adopted the Rose-Brink phase convention. The value listed here has been converted to the Steffen-Krane convention.} 
\footnotetext[5]{From ENSDF \cite{ensdf2021_11}. For the first two transitions, the values originate from Ref.~\cite{PhysRev.139.B512} and do not sum to 100\%.} 
\end{center}
\end{table*}
%

\subsection{Transitions of the $8920$-keV level to and through the $4445$-keV state}
\begin{figure}
\includegraphics[width=1.0\columnwidth]{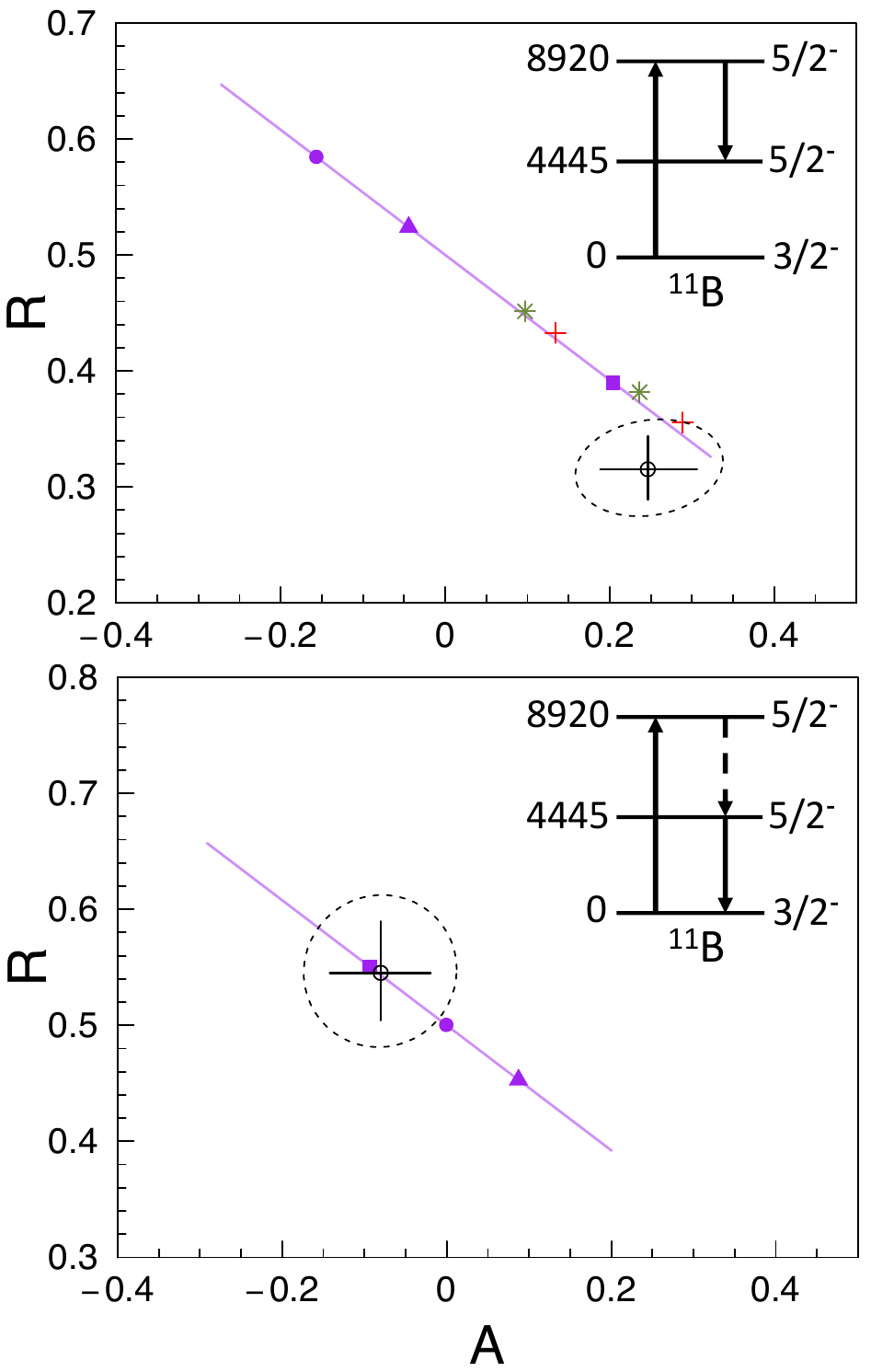}
\caption{\label{fig:b114445} 
Ratio {\it vs.} analyzing power for the deexcitation of the $8920$-keV ($5/2^-$) level in $^{11}$B. Measured values are depicted in black, where the dashed black lines indicate 68\% error ellipses. The purple lines depict the theoretical solutions for a range of mixing ratios. Purple circles, squares, and triangles indicate theoretical values of $\delta_2$ $=$ $-10$, $0$, and $+10$, respectively. (Top) Primary $8920$~keV $\rightarrow$ $4445$~keV transition. The purple line was obtained with the present value of the mixing ratio for the transition exciting the $8920$-keV level (see Tab.~\ref{tab:11b_delta} and Sec.~\ref{sec:8920to0}) and a range of mixing ratios, $\delta_2$, for the primary decay to the $4445$-keV state. The region between the green asterisks or red crosses corresponds to the two possible solutions from our reanalysis of the data by Ref.~\cite{Rusev:2009eo} (for $\delta_1$ $=$ $-0.001$; see App.~\ref{sec:app} and Tab.~\ref{tab:app}). (Bottom) Secondary $4445$~keV $\rightarrow$ $0$~keV transition, with the primary transition unobserved. The purple line was obtained assuming the present value of the mixing ratio for the transition exciting the $8920$-keV level, a mixing ratio of $|\delta_u|$ $=$ $0.72$ for the unobserved intermediate transition (see Tab.~\ref{tab:11b_delta}), and a range of mixing ratios, $\delta_2$, for the secondary decay of the $4445$-keV state.
}
\end{figure}

Our experimental results for the $8920$~keV $\rightarrow$ $4445$~keV transition are presented as a black data point ($A$ $=$ $+0.25\pm0.06$ and $R$ $=$ $0.32\pm0.03$) and $68$\% error ellipse in the top panel of Fig.~\ref{fig:b114445}. The purple line was calculated with the present value of the mixing ratio, $\delta_1$, for the transition exciting the $8920$-keV level (Sec.~\ref{sec:8920to0}) and a range of mixing ratios, $\delta_2$ $=$ $-10$ to $+10$, for the primary decay to the $4445$-keV state. 
The purple symbols have the same meaning as in Fig.~\ref{fig:b11AC}. A Bayesian fit to our data yields a value for the mixing ratio of $\delta_2^{present}=$ $+0.72 \pm 0.40$. In comparison, the reanalysis of the Rusev {\it et. al.} \cite{Rusev:2009eo} data, and adopting just one of their four solutions for the ground-state transition ($\delta_1$ $=$ $-0001$), yields two solutions for the transition to the $4445$-keV state: $\delta_2^{Rusev}$ $=$ $-0.08^{+0.13}_{-0.14}$ and $+1.7^{+0.7}_{-0.6}$ (Tab.~\ref{tab:app}). These are marked in the top panel of Fig.~\ref{fig:b114445} by pairs of green asterisks and red crosses, respectively. It can be seen that our error ellipse (dashed black line) covers a significantly more constrained range of mixing ratios, when compared to the results of Ref.~\cite{Rusev:2009eo}.

As a final test of the internal consistency of our data, we present in the lower panel of Fig.~\ref{fig:b114445} our experimental results for the $4445$~keV $\rightarrow$ $0$~keV secondary transition ($A$ $=$ $-0.08\pm0.06$ and $R$ $=$ $0.55\pm0.05$). In this case, the purple line was computed assuming the present value of the mixing ratio for the transition exciting the $8920$-keV level (see Tab.~\ref{tab:11b_delta} and Sec.~\ref{sec:8920to0}), a mixing ratio of $|\delta_u|$ $=$ $0.72$ for the unobserved intermediate radiation, and a range of mixing ratios, $\delta_2$, for the secondary decay of the $4445$-keV state. Note that this type of angular correlation is independent of the sign of the mixing ratio for the unobserved intermediate transition, $\delta_u$ \cite{iliadis21}. It can be seen that the present data are not sufficiently precise to deduce a meaningful mixing ratio, $\delta_2$, for the $4445$~keV $\rightarrow$ $0$~keV transition. This is not surprising, because the transition appears in our measurement as the last step in a three-photon process. All that can be concluded from the lower panel of Fig.~\ref{fig:b114445} is that the present data are consistent with a broad range of mixing ratios, $\delta_2$ once we fix the mixing ratios, $\delta_1$ and $\delta_u$, for the preceding steps using our measured values (Tab.~\ref{tab:11b_delta}).

Bell {\it et. al.} \cite{BELL1968481} utilized the $^{12}$C(t,$\alpha$)$^{11}$B reaction to measure the angular correlation of the $4445$~keV $\rightarrow$ $0$~keV transition, resulting in a value of $\delta^{\text{Bell}}$ $=$ $+0.19 \pm 0.03$. Rusev {\it et. al.} \cite{Rusev:2009eo} excited the $4445$-keV level by NRF and the reanalysis of their data yields four solutions (see App.~\ref{sec:app}): $\delta^{\text{Rusev}}$ $=$ $-2.90^{+0.13}_{-0.18}$, $-0.511^{+0.011}_{-0.017}$, $+0.172^{+0.019}_{-0.020}$, and $+2.13^{+0.06}_{-0.06}$. The third solution agrees with the result of Ref.~\cite{BELL1968481}. Although it is reasonable to exclude the first and fourth solutions, based on an implied unusually large $E2$ admixture, it would be questionable to disregard the third solution based on the evidence from the data by Ref.~\cite{Rusev:2009eo} alone. In view of the poor detector resolution in the experiment of Bell {\it et al.} \cite{BELL1968481}, and the nearly isotropic angular correlation shown in their Fig.~5, a remeasurement of the $4445$~keV $\rightarrow$ $0$~keV angular correlation in an NRF experiment by using several detectors placed inside and outside the $\theta$ $=$ $90^\circ$ plane is recommended.

\subsection{Branching ratios for the $8920$-keV ($5/2^-$) level}
Gamma-ray branching ratios for the $8920$-keV state were derived from our angle-integrated yields. We find branching ratios of $B_\gamma$ $=$ $96.83\pm0.13$\% and $3.17\pm0.13$\% for the $8920$~keV $\rightarrow$ $0$~keV and $8920$~keV $\rightarrow$ $4445$~keV transitions, respectively (see Tab.~\ref{tab:11b_delta}). Our values are more precise than those listed in ENSDF \cite{ensdf2021_11}. Notice that the erroneous relationship for the mixing ratios assumed in Ref.~\cite{Rusev:2009eo} also impacts their reported $\gamma$-ray branching ratios. However, insufficient information is provided in their work for us to derive corrected values.

\section{Summary}\label{sec:summary}

We report on a measurement of $^{11}$B and $^{40}$Ca levels at energies between $8$ and $9$~MeV using nuclear resonance fluorescence (NRF) at the High-Intensity $\gamma$-ray Source (HI$\gamma$S) facility at the Triangle Universities Nuclear Laboratory (TUNL). Levels in $^{11}$B are interesting because they provide calibration points for NRF experiments, while states in the $^{40}$Ca nucleus represent low-energy resonances in the $^{39}$K(p,$\gamma$)$^{40}$Ca reaction, which is of relevance for potassium nucleosynthesis in globular clusters. 

We present improved excitation energies for $^{40}$Ca as well as an unambiguous $2^-$ assignment for the state at $8425$~keV. We also obtained improved $\gamma$-ray multipolarity mixing ratios and more precise branching ratios for the $^{11}$B state at $8920$~keV. Furthermore, previously published values of mixing ratios in $^{11}$B by Rusev {\it et al.} \cite{Rusev:2009eo} were derived under an erroneous assumption concerning phase conventions in angular correlation expressions. We reanalyzed their data and derived corrected values for the mixing ratios in $^{11}$B. We also comment on the relationship of different phase conventions in NRF experiments.


\begin{acknowledgments}
We would like to thank Gencho Rusev for detailed discussions about the mixing ratio analysis presented in Ref.~\cite{Rusev:2009eo}, and Richard Longland, Samantha Johnson, and S. Gaither Frye for providing comments. The help of Mark Emamian with preparing Fig.~1 is highly appreciated. This work is supported by the DOE, Office of Science, Office of Nuclear Physics, under grants DE-FG02-97ER41041 (UNC) and DE-FG02-97ER41033 (TUNL). 
\end{acknowledgments}

\appendix
\section{Note on phase conventions}\label{sec:phase}
When multipolarity mixing ratios are involved in transitions, it is important to clearly state the adopted phase conventions. We analyzed our data using the expressions presented in Ref.~\cite{iliadis21}, which are based on the work of Biedenharn \cite{biedenharn60}. In the Biedenharn phase convention, the intermediate state always appears on the right in the reduced matrix elements, regardless of whether it is the initial or final state involved in the transition. Applied to an NRF experiment on an odd-mass target nucleus, where the excited state decays back to the ground state (i.e., elastic $\gamma$-ray scattering), this implies a relation of $\delta_1$ $=$ $\delta_2$ for the mixing ratios of the first (excitation) and second (deexcitation) step, as pointed out in Sec.~3 of Ref.~\cite{iliadis21}.  

The original formulation of the Steffen-Krane formalism \cite{Krane:1973wr,kranesteffen} provides the expressions for $\gamma$-ray cascades. In their phase convention, the initial state always appears on the right in the reduced matrix elements. Applying these expressions directly to an NRF experiment on an odd-mass target nucleus, where the excited state decays back to the ground state (see, e.g., Eqs.~(4) $-$ (7) in Ref.~\cite{Pietralla03}), implies a relation of $\delta_1$ $=$ $-\delta_2$ for the mixing ratios, as pointed out by Refs.~\cite{Martin:1987uu,iliadis21}. If one modifies the original expressions of the Steffen-Krane formalism \cite{Krane:1973wr,kranesteffen} by switching the sign in front of the terms containing $\delta_1$ (see, e.g., Refs.~\cite{Kneissl1996,zilges2022}), then the mixing ratios are related by $\delta_1$ $=$ $\delta_2$. The error made in Ref.~\cite{Rusev:2009eo} was the assumption of $\delta_1$ $=$ $\delta_2$, while adopting at the same time the original \cite{Krane:1973wr,kranesteffen} instead of the modified Steffen-Krane expressions. We present a reanalysis of the previous data in App.~\ref{sec:app}. 

A final note regarding phase conventions: for the processes discussed here, i.e., elastic or inelastic $\gamma$-ray scattering, and an unobserved intermediate transition in a two-photon deexcitation, the formalisms of Biedenharn \cite{biedenharn60} and Steffen-Krane \cite{Krane:1973wr,kranesteffen} (using either their original or modified expressions) provide {\it the same sign} for the mixing ratio of the second (deexcitation) transition. To avoid any confusion, we quote in the present work mixing ratio values for deexcitation transitions only.

\section{Reanalysis of the Rusev {\it et al.} \cite{Rusev:2009eo} data}\label{sec:app}
Here, we describe our reanalysis of the data of Ref.~\cite{Rusev:2009eo} and present the results. In the original publication, the authors used an erroneous phase convention for the multipolarity mixing ratio, as described in App.~\ref{sec:phase}.

Rusev \textit{et al.} \cite{Rusev:2009eo} measured the asymmetry, $a$. Since their definition of the azimuthal angle, $\phi$, differs from our convention (see Eq.~(\ref{eq:analyzing}) and Ref.~\cite{iliadis21}), their asymmetry is related to our analyzing power by $a$ $=$ $- A(\theta = 90^\circ)$. Their method of taking into account geometric effects in the quantity $a$ did not use $Q$ coefficients, but relied on separate {\sc Geant3} simulations for each cascade of interest. While not mentioned explicitly, they assumed that their geometry-correction factors, $C \left( \theta \right)$,  are independent of the mixing ratio. Rusev \textit{et al.} \cite{Rusev:2009eo} observed  ground-state and excited-state transitions (which they call ``branching transitions''), with asymmetries of $a_{gs} \left( \delta_1 \right)$ and $a_{ex} \left( \delta_1, \delta_2 \right)$, respectively. The data were analyzed in a two-step process. The first step used only the ground-state transition and employed the experimental mode, $\bar{a}_{gs}$, and the standard deviation, $\Delta a_{gs}$, to find a mode $\bar{\delta}_1$ and a (potentially asymmetric) coverage interval $\left[ \bar{\delta}_1 - \Delta_- \delta_1, \bar{\delta}_1 + \Delta_+ \delta_1 \right]$ for the mixing ratio,
\begin{align}
    \label{eq:min_max}
    \bar{\delta}_1 &= a_{gs}^{-1} \left( \bar{a}_{gs} \right) \\
    \bar{\delta}_1 \pm \Delta_\pm {\delta}_1 &= a_{gs}^{-1} \left( \bar{a}_{gs} \pm \Delta a_{gs} \right). \nonumber
\end{align}
The inverse of $a_{gs}$ in this so-called ``min-max'' method, $a_{gs}^{-1}$, was determined numerically. The result can be visualized as a projection of the experimental coverage interval for $a_{gs}$ on the mixing-ratio axis. 

\begin{figure}
\includegraphics[width=1.0\columnwidth]{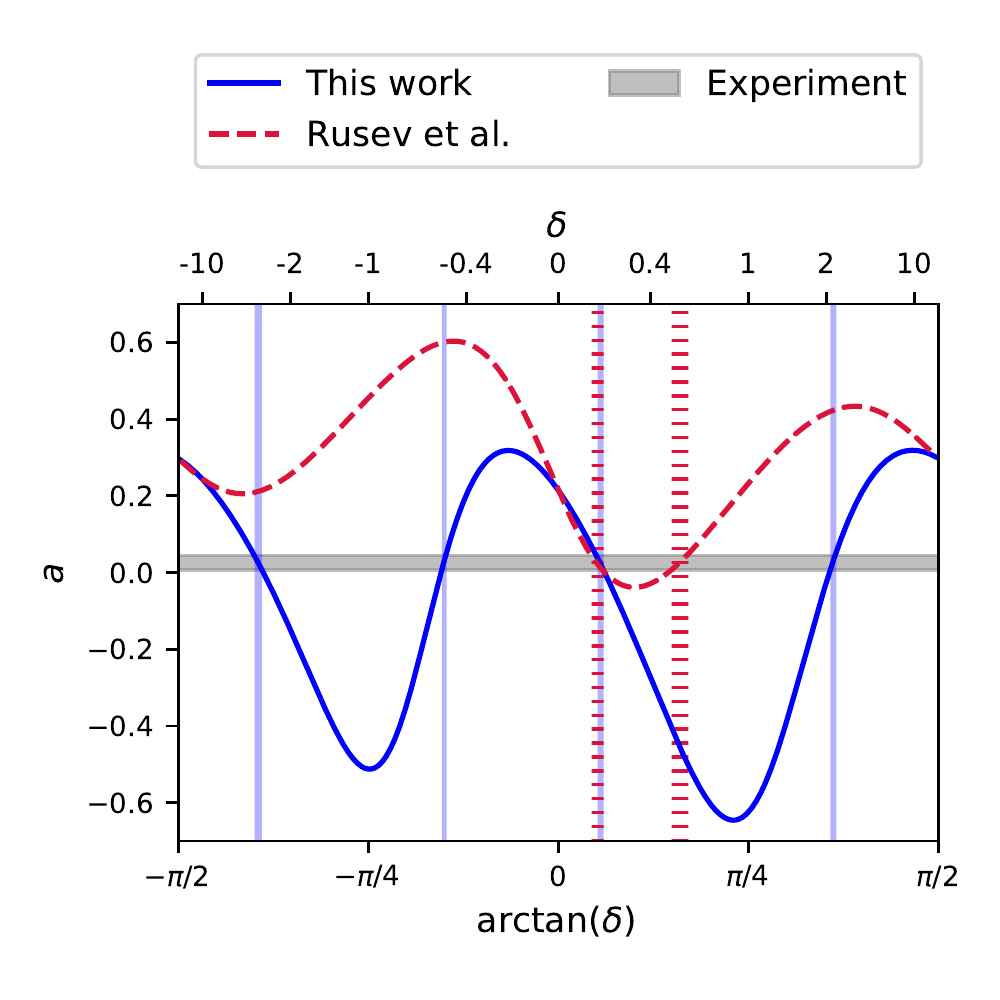}
\caption{\label{fig:projection_method} 
Determination of the multipolarity mixing ratio, $\delta$, for the sequence $0$~keV ($3/2^-$) $\rightarrow$ 4445~keV ($5/2^-$) $\rightarrow$ $0$~keV ($3/2^-$) in $^{11}$B from the experimental asymmetry, $a$ $=$ $- A(\theta = 90^\circ)$, using the ``min-max method.'' The solid blue curve indicates the theoretical dependence of the asymmetry on the mixing ratio. The dashed red curve is instead obtained, if the erroneous mixing-ratio phase convention of Ref.~\cite{Rusev:2009eo} is adopted. Both curves take into account the correction for the finite opening angle of the detectors (factor $0.95$ \cite{Rusev:2009eo}). The asymmetry measured by Ref.~\cite{Rusev:2009eo} is shown as a solid gray horizontal region. The possible solutions for the multipolarity mixing ratio, $\delta$, are depicted by vertical bars (solid blue and dashed red for the correct and erroneous phase convention, respectively).  
}
\end{figure}

As an example, the determination of $\delta$ for the sequence $0$~keV ($3/2^-$) $\rightarrow$ 4445~keV ($5/2^-$) $\rightarrow$ $0$~keV ($3/2^-$) in $^{11}$B is given in Fig.~\ref{fig:projection_method}. The solid blue curve indicates the theoretical dependence of the asymmetry on the mixing ratio using the correct phase convention (see App.~\ref{sec:phase}). The dashed red one is obtained instead, if the erroneous mixing-ratio phase convention of Ref.~\cite{Rusev:2009eo} is adopted\footnote{Rusev et al. \cite{Rusev:2009eo} show the dependence of their asymmetry on the mixing ratio for the sequence $0$~keV ($3/2^-$) $\rightarrow$ 4445~keV ($5/2^-$) $\rightarrow$ $0$~keV ($3/2^-$) in $^{11}$B as a black curve in their Fig.~4. Their dependence disagrees with our dashed red curve in Fig.~\ref{fig:projection_method}, although both curves were obtained using the erroneous phase convention (App.~\ref{sec:phase}). We can reproduce the black curve in Fig.~4 of Ref.~\cite{Rusev:2009eo} if, in addition to an erroneous phase convention, we also adopt an erroneous sign in front of the $P^{|2|}_4$ term in Eq.~\eqref{eq:angcor}. The $P^{|2|}_4$ term is proportional to $\delta^4$ for this sequence (see Ref.~ \cite{zilges2022}), which would explain why an erroneous sign affects mainly large values of the mixing ratio, $\delta$.}. The experimental asymmetry of Ref.~\cite{Rusev:2009eo} is indicated as a gray horizontal region. The possible solutions for the multipolarity mixing ratio, $\delta$, are depicted by vertical bars (solid blue strip and red dashes for the correct and erroneous phase convention, respectively). The significant difference in the theoretical descriptions is apparent.

In a second step, Ref.~\cite{Rusev:2009eo} employed the estimated value of $\delta_1$ to determine the mixing ratio, $\delta_2$, for the transition to an excited level, again using the ``min-max method.'' One shortcoming of this approach is the implicit assumption that $a_{gs}$ or $a_{ex}$ increases or decreases monotonously with the mixing ratio. However, the dependence of the asymmetry on a mixing ratio results frequently in multiple local extrema (see below). Furthermore, Ref.~\cite{Rusev:2009eo} did not propagate the uncertainties from the first step to the second one.

In our reanalysis of their data, we improved the ``min-max method'' by propagating the uncertainty using the Monte Carlo method described in Ref.~\cite{gum08}. In a first step, a random value of $a_{gs}^\mathrm{rand}$ was sampled from a normal distribution, $\mathcal{N} \left( \bar{a}_{gs}, \Delta a_{gs} \right)$. By numerical inversion, $a_{gs}^{-1} \left( a_{gs}^\mathrm{rand} \right)$, a set of solutions $\left\{ \delta_{1}^\mathrm{rand} \right\}$ was found. In a second step, the inversion $a_{ex}^{-1} \left( \delta_1^\mathrm{rand}, a_{ex}^\mathrm{rand} \right)$, which used the set of randomly-sampled values $\delta_1^\mathrm{rand}$, was used to find a set of solutions $\left\{\delta_2^\mathrm{rand} \right\}$. From a total of $10^5$ valid samples for each excited state (i.e., samples for which the inversion is possible for all observed transitions), probability distributions for all mixing ratios can be estimated. These distributions naturally take correlations into account, and are not subject to the drawbacks of the min-max method.

Our results are presented in Tab.~\ref{tab:app}. When the sampling yielded unimodal probability densities for a mixing ratio, we quote the \unit[68]{\%} highest-density interval. In cases where the distributions encompassed unbounded regions of parameter space, we provide lower or upper limits. If Tab.~\ref{tab:app} shows the word ``undetermined'' for a mixing ratio of the transition to a low-lying excited state, this means that all ranges were approximately equally probable.  It should be noted that we were able to reproduce all the multipolarity-mixing-ratio results reported by Ref.~\cite{Rusev:2009eo} when their erroneous phase convention is used.

%
\begin{table*}
\begin{center}
\caption{Mixing ratios (consistent with the Steffen-Krane phase convention) in $^{11}$B from our reanalysis of the Rusev {\it et al.} \cite{Rusev:2009eo} data.}
\label{tab:app}
\begin{ruledtabular}
\begin{tabular}{lcclc}
Transition (keV)\footnotemark[1] &    $E_\gamma$ (keV)\footnotemark[2] & $a$\footnotemark[3] & $\delta_1$\footnotemark[4] & $\delta_2$\footnotemark[4] \\
       \hline
4445 ($5/2^-$) $\rightarrow$ 0 ($3/2^-$)        &    4444.03(8)     &  $+$0.026(19)   & $-2.90^{+0.13}_{-0.18}$    \\
                                                &                   &               & $-0.511^{+0.011}_{-0.017}$ &  \\
                                                &                   &               & $+0.172 \pm 0.020$ &  \\
                                                &                   &               & $+2.13 \pm 0.06$ &  \\
\hline
5020 ($3/2^-$)  $\rightarrow$ 0 ($3/2^-$)       &    5018.98(40)    &  $+$0.200(16)   & $-0.055 \pm 0.023$ &  \\
5020 ($3/2^-$) $\rightarrow$ 2125 ($1/2^-$)     &    2895.30(40)    &  $-$0.37(5)     & $-0.055 \pm 0.023$ & $-1.39^{+0.18}_{-0.22}$ \\
                                                &                   &               & $-0.055 \pm 0.023$ & $-0.08^{+0.06}_{-0.08}$ \\
\hline
7286 ($5/2^+$) $\rightarrow$ 0 ($3/2^-$)        &    7282.92        &  $-$0.213(30)   & $< -5$    \\
                                                &                   &               & $-0.380^{+0.030}_{-0.028}$ &  \\
                                                &                   &               & $+0.003 \pm 0.038$ &  \\
                                                &                   &               & $+3.1^{+0.4}_{-0.6}$ &  \\
7286 ($5/2^+$) $\rightarrow$ 5020 ($3/2^-$)     &    2264.9         &  $-$0.181(80)   & $< -5$ & $< -2$ \\
                                                &                   &               &  & $+ 0.40^{+0.15}_{-0.10}$ \\
                                                &                   &               &  & $> + 1$ \\
                                                &                   &               & $-0.380^{+0.030}_{-0.028}$ & $-1.8^{+0.8}_{-1.6}$ \\
                                                &                   &               &  & $-0.25^{+0.15}_{-0.18}$ \\
                                                &                   &               & $+0.003 \pm 0.038$ & $< -2.1$ \\
                                                &                   &               &  & $+0.03 \pm 0.08$ \\
                                                &                   &               & $+3.1^{+0.4}_{-0.6}$ & undetermined \\
7286 ($5/2^+$) $\rightarrow$ 4445 ($5/2^-$)     &    2840.23        &  $+$0.177(66)   & $< -5$ & $-0.70^{+0.15}_{-0.23}$ \\
                                                &                   &               &  & $> +1.7$ \\
                                                &                   &               & $-0.380^{+0.030}_{-0.028}$ & undetermined \\
                                                &                   &               & $+0.003 \pm 0.038$ & $-0.10 \pm 0.13$ \\
                                                &                   &               &  & $ > +1.1$ \\
                                                &                   &               & $+3.1^{+0.4}_{-0.6}$ & undetermined \\
\hline
8920 ($5/2^-$) $\rightarrow$ 0 ($3/2^-$)        &    8916.67(16)    &  $+$0.215(20)   & $-6.7^{+0.9}_{-1.8}$ & \\
                                                &                   &               & $-0.387 \pm 0.019$ & \\
                                                &                   &               & $-0.001 \pm 0.028$ & \\
                                                &                   &               & $+3.06^{+0.28}_{-0.19}$  & \\
8920 ($5/2^-$) $\rightarrow$ 4445 ($5/2^-$)     &    4474.5(3)      &  $-$0.19(7) \footnotemark[10] & $-6.7^{+0.9}_{-1.8}$ & $< -2.1$   \\
                                                &                   &               &  & $-0.72^{+0.16}_{-0.24}$ \\
                                                &                   &               &  & $> + 1.7$ \\
                                                &                   &               & $-0.387 \pm 0.019$ & undetermined \\
                                                &                   &               & $-0.001 \pm 0.028$ & $-0.08 \pm 0.14$ \\
                                                &                   &               &  & $+1.7 \pm 0.7$ \\
                                                &                   &               & $+3.06^{+0.28}_{-0.19}$ & undetermined \\
\end{tabular}
\end{ruledtabular}
\footnotetext[1]{Deexcitation transition of the cascade under consideration, with the energy of the excited states given in keV, and spin-parity values in parentheses; the  excitation is always from the $3/2^-$ ground state of $^{11}$B.}
\footnotetext[2]{Gamma-ray energy adopted from Ref.~\cite{ensdf2021_11}.}
\footnotetext[3]{Experimental asymmetry from Ref.~\cite{Rusev:2009eo}, which is related to our analyzing power by $a$ $=$ $- A(\theta = 90^\circ)$.}
\footnotetext[4]{Multipolarity mixing ratios from our reanalysis of the Rusev et al. data \cite{Rusev:2009eo}. For ground-state transitions, the mixing ratio is the same for the excitation and deexcitation transitions of the cascade (see App.~\ref{sec:phase}) and, therefore, only $\delta_1$ is given. Multiple entries for the same transition correspond to alternative solutions.}
\footnotetext[10]{Uncertainty reported in the literature has a typo and is a factor of 10 too small \cite{Rusev_PrivCom}
.}
\end{center}
\end{table*}

\section{Energy calibration procedure}\label{sec:ener}
Four precisely known $\gamma$-ray energies of transitions in $^{11}$B were used for the detector energy calibrations: $E_{\gamma}$ $=$ $4474.51(13)$, $4666.05(30)$, $6789.55(30)$, and $8916.59(11)$~keV. These values were obtained from the recommended excitation energies listed in Ref.~\cite{ensdf2021_11}, and corrected for recoil shifts. Peak centroids and uncertainties (in channel units) were estimated using ``Method C'' of Ref.~\cite{RODGERS2021165172}.

A linear energy calibration was then performed using a Bayesian regression model. We used three independent chains with 500\,000 steps each (with a burn-in of 50\,000 steps), to ensure convergence of the Markov chains. One novel aspect of our fitting procedure is the inclusion of uncertainties in both $x$ (channel centroid) and $y$ ($\gamma$-ray calibration energy), which is rigorously (in a statistical sense) implemented using the Bayesian model described in Sec.~8.4 of Ref.~\cite{andreonWeaver}. The same technique was also successfully applied in Ref.~\cite{deSouza:2019gf} in connection with the fitting of astrophysical S-factors. 

\begin{figure}
\includegraphics[width=1.0\columnwidth]{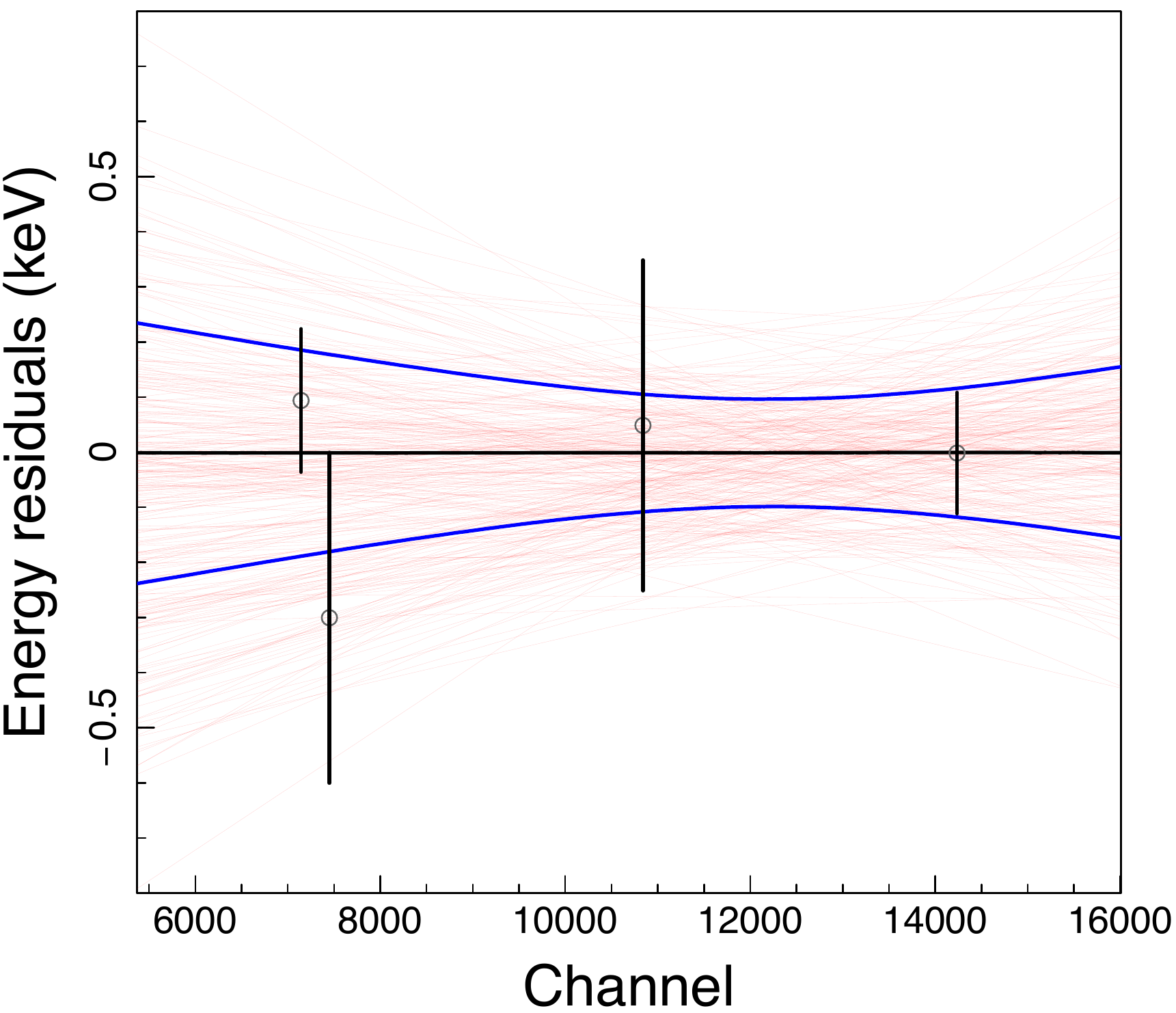}
\caption{\label{fig:enercal} 
Energy residuals (i.e., the difference with respect to the best fit line) versus peak centroid (in channels) for detector 2 (D2; see Fig.~\ref{fig:setup}). The four data points refer to the four $^{11}$B calibration energies (see text). The red lines show credible lines for different steps of the Markov chains. The two blue lines enclose a coverage probability of 68\%. The region of interest for the $^{40}$Ca levels measured in the present work is located between channels 13\,000 to 14\,000. Only 300 out of 50\,000 regression lines are plotted for the purpose of illustration.
}
\end{figure}

Figure~\ref{fig:enercal} displays the energy residuals (i.e., the difference with respect to the best fit line) versus peak centroid (in channels) for detector 2 (D2; see Fig.~\ref{fig:setup}). The four data points refer to the four $^{11}$B calibration energies quoted above. The red lines show, in the parlance of Bayesian statistics, ``credible lines'' for different steps of the Markov chains. The 16 and 84 percentiles of the energy residuals are plotted as blue lines. It can be seen that, for a coverage probability of 68\%, the uncertainty resulting from the linear fit is $\approx 0.1$~keV near the $\gamma$-ray energy region of interest ($8$ $-$ $9$~MeV, corresponding to channels 13\,000 to 14\,000).

To this (common) uncertainty, one must add that resulting from the measured $^{40}$Ca peak centroids (in channels). This is achieved by predicting a value, $y_p$, of the $\gamma$-ray energy according to (see, e.g., Sec.~8.10 in Ref.~\cite{andreonWeaver})
\begin{equation}
   p(y_p | x_p, D) = \int_\theta p(y_p | x_p , \theta) p(\theta | D) d\theta 
\end{equation}
where $x_p$ is the experimentally determined peak centroid, $\theta$ denotes the vector of model parameters (slope and intercept of the linear model), $D$ stands for the set of data, and $p(\theta | D)$ is the posterior (i.e., the probability density for obtaining parameter values of $\theta$ given the data, $D$). To account for the experimentally-determined uncertainty of a $^{40}$Ca peak centroid, we sampled $x_p$ from a Gaussian probability density with a mean and standard deviation given by the peak centroid and its uncertainty, respectively. Finally, the $\gamma$-ray energies and uncertainties for observed decays in $^{40}$Ca (see Tab.~\ref{tab:40ca_energies}) are derived from the 50 percentile (median) and the 16 and 84 percentiles (68\% coverage probability) of the probability density, $p(y_p | x_p, D)$.

We did not include the single-escape or double-escape peaks corresponding to the $8916.59(11)$~keV full-energy peak from $^{11}$B in our linear energy calibration, although these three peaks together cover the entire energy range of all $^{40}$Ca ground-state decays observed in this work. Previous work \cite{KIKSTRA1990425} has shown that the energy differences between escape peaks and the full energy peak are systematically shifted from $511$ or $1022$~keV. These shifts depend on the type of the detector and amount to about $100$~eV for the hyperpure n-type germanium detector used in Ref.~\cite{KIKSTRA1990425}. Since this shift is very small, we performed a test by including the escape peaks in our linear energy calibration. As a result, the $\gamma$-ray energies of all transitions given in column 3 of Tab.~\ref{tab:40ca_energies} varied by less than their stated standard deviation.

Potential amplifier gain shifts were monitored throughout the experiment using room background peaks (with beam on the sample) and $\gamma$-ray lines from radioactive sources (without beam), and no gain shifts were observed.

\bibliography{paper}

\end{document}